\documentclass[fleqn,twoside]{article}
\usepackage{espcrc2}

\usepackage{graphicx}
\usepackage[figuresright]{rotating}

\newcommand{\AmS}{{\protect\the\textfont2
  A\kern-.1667em\lower.5ex\hbox{M}\kern-.125emS}}

\title{Testing new strategies in finite density}

\author{
V. Azcoiti \address[zara]{Departamento de F\'{\i}sica Te\'orica, 
Universidad de Zaragoza, E-50009 Zaragoza, Spain},
G. Di Carlo\address[lngs]{INFN, Laboratori Nazionali del Gran Sasso, 
67010 Assergi, (L'Aquila) Italy},
A. Galante\addressmark[lngs]\address[aq]{Dipartimento di Fisica 
dell'Universit\`a di L'Aquila, 67100 L'Aquila, Italy}
        \thanks{Talk presented by A. Galante.},
V. Laliena\addressmark[zara].
}

\begin{document}

\begin{abstract}
A new approach for non zero chemical potential simulations is
tested in the Gross-Neveu model for infinite flavor number,
where the critical line is
reconstructed in a large $\mu/T$ interval. A comparison
with results from standard imaginary chemical potential 
approach as well as first results for $N_f=4$ QCD are 
presented.

\vspace{1pc}
\end{abstract}

\maketitle

Recently we proposed a new method for simulate systems at non
zero chemical potential \cite{noi}. The starting point is to note that the
temporal part of the Dirac operator (the one that contains the 
$\mu$ dependence) can be written as the sum of two contributions:
the first, proportional to $\cosh(\mu)$, multiplies an antihermitian
operator, the second, proportional to $\sinh(\mu)$, multiplies
an hermitian operator. 
Since the spacial part of the fermionic operator is antihermitian
too, we see that for non zero $\mu$ the Dirac matrix looses its
global antihermiticity and this is the origin of the complex
action problem.

Clearly the imaginary chemical potential solves the problem:
$\sinh(\mu)$ gets an extra factor
$i$ and converts the corresponding operator to an antiermitian one.
Introducing an imaginary chemical potential we look 
for a critical line in the $(\beta,i\mu)$ plane to analytically 
continue in the $(\beta,\mu)$ one: in doing that we are forced 
to consider larger values of the physical temperature $T$ than 
the ones of physical relevance.

In our proposal we promote $\cosh(\mu)$ and $\sinh(\mu)$ to independent
variables $x$ and $y$ and explore the phase structure as a function
of such variables to recover the physical critical point 
along the line that corresponds to $x=\cosh(\mu)$, $y=\sinh(\mu)$.
To do that we perform standard ($i.e.$ with real and positive action) 
simulations at pure imaginary $y$ and extend the results to real
$y$.
This can be achieved in two different ways \cite{noi}. 

The first is to fix
the inverse gauge coupling and $(i)$ find the critical line in the
$(x,iy)$ plane, $(ii)$ fit this line with an even polynomial 
of $y$, $(iii)$ analytically continue the critical line to the 
$(x,y)$ plane, $(iv)$ from the intersection of the analytically continued
critical line with the $x^2-y^2=1$ line (which corresponds to 
real and positive values of $\mu$) it is possible to read the 
critical value of $\mu$. Working at fixed lattice temporal extent
$(N_t)$ this is equivalent to do simulations at fixed $T$ $i.e.$
the same physical temperature of the physical critical point.

The second approach is to choose a value for $\mu$ and fix 
the $x$ variable to $\cosh(\mu)$. By standard simulations we
find the critical line in the $(\beta,iy)$ plane and proceed
in close analogy with previous case. Only point $(iv)$
changes: to determine the critical
value of $\beta$ at given $\mu$ we look for 
the intersection of the 
analytically continued critical line with 
$y=\sinh(\mu)$.
In this case, for a given
$N_t$, this is equivalent to search for the transition
point doing simulations at fixed $\mu/T$.

\begin{figure}[t]
\vspace{9pt}
\centerline{\includegraphics*[width=2.3in,angle=270]{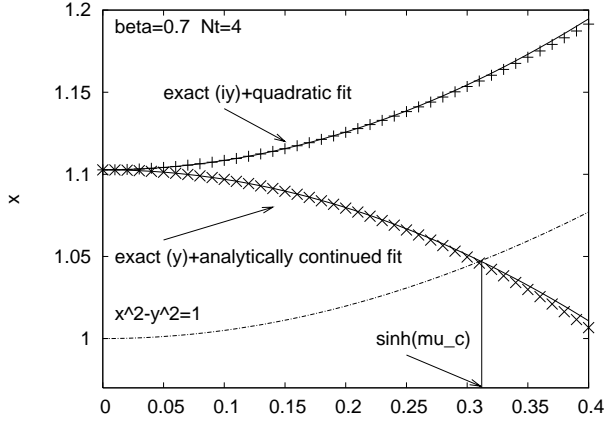}}
\caption{G-N model: analytical continuation of the critical line from
imaginary to real $y$ and determination of the critical $\mu$.}
\label{fig1}
\end{figure}
Comparing the standard imaginary chemical potential
approach \cite{phil} and our approach(es) we can
note that in both case we have to rely on an analytical 
continuation of the critical line (with all the uncontrolled 
systematics that this procedure can in principle introduce).

The simulation cost of our method(s) is bigger 
since we need to explore a two dimensional space (the $x-iy$ or
$\beta-iy$ one) to extract the critical line.
Besides that we can expect to be somehow closer to the
physical critical point (in one case we work at same $\beta$,
in the other at the same value for the adimensional lattice 
$\mu$) and this could improve to reduce the overall systematic 
effects due to extrapolation ($i.e.$ reach larger $\mu/T$ values).

To check if the above prejudices are correct or not we
consider an analytically solvable model:
the Gross-Neveu model ($i.e.$ the four-Fermi model
in $2+1$ dimensions) at finite $T$ and $\mu$.

\begin{figure}[t]
\vspace{9pt}
\centerline{\includegraphics*[width=2.3in,angle=270]{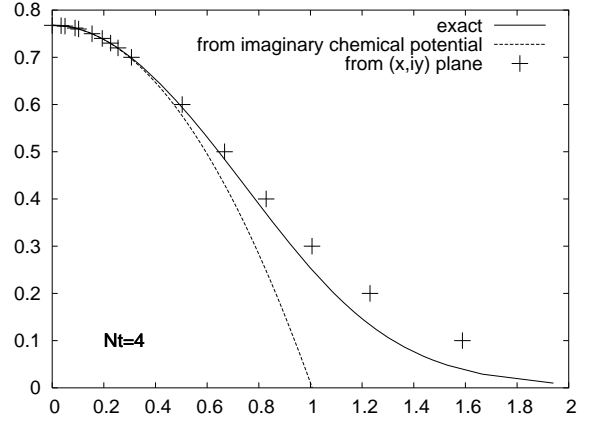}}
\caption{G-N model: the critical line $\beta(\mu)$.}
\label{fig2}
\end{figure}
It can be solved analytically in the large $N_f$ limit 
and its phase structure in the $(\mu, T)$ plane resembles very
much the one expected for QCD, with a phase transition line separating a
phase where chiral symmetry is spontaneously broken from a phase where
the symmetry is restored \cite{hands}.

After the introduction of an auxiliary scalar field $\sigma$
we can define a generalized fermion action (we consider the zero
mass limit) 
\begin{eqnarray}
S_F  \left(  x,y\right)= S_F^{sp}+{1\over 2}\sum^{N_f/2}_{i=1}\sum_{n}
\Bigg\{ 
    x \bar\psi^i_{n}
    \left[
      \psi^{(i)}_{n+0} - \psi^{(i)}_{n-0}
    \right]
 \nonumber\\
+  y \bar\psi^{(i)}_{n}
  \left[
    \psi^{(i)}_{n+0} + \psi^{(i)}_{n-0}
  \right]
+ {1\over 4}
\bar\psi^{(i)}_n\psi^{(i)}_n \sum_{<\bar n, n>} \sigma_{\bar n}
\Bigg\}\nonumber
\label{grossneveu}
\end{eqnarray}
\noindent
where $S_F^{sp}$ is the standard spacial part of the Kogut-Susskind
action and $<\bar n, n>$ denotes the set of 8 dual lattice sites $\bar n$
surrounding the site $n$. 
\begin{figure}[t]
\vspace{9pt}
\centerline{\includegraphics*[width=2.3in,angle=270]{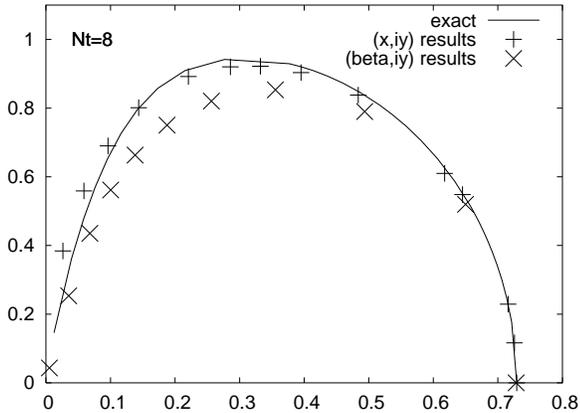}}
\caption{G-N model: comparison of our results from $(x,iy)$ and $(\beta,iy)$
analysis and exact result in the $\mu/\Sigma_0-T/\Sigma_0$ plane.}
\label{fig3}
\end{figure}
The above actions agrees with the 
standard one for  $x=\cosh(\mu)$, $y=\sinh(\mu)$. 
In the $N_f\to\infty$ limit the generalized model can be analytically 
solved by the gap equation 
\begin{equation}
\Sigma = -g^2\left\langle \bar\psi \psi\right\rangle \nonumber
\label{gap}
\end{equation}
where $\Sigma$ is the expectation value of the scalar field.

We used (\ref{gap}) in the infinite spacial volume limit
to find the critical line in the $(x,iy)$ and
$(\beta,iy)$ planes and then apply our procedures to extract the
critical line as a function of $y$.
These results can be directly compared with the exact ones we can
extract from the gap equation in order to check, in a non trivial case,
the potentialities of our approach.
The final goal is to reconstruct the critical line in the $(\beta,\mu)$
plane or, using adimensional variables, in the 
$(\mu/\Sigma_0,T/\Sigma_0)$ one where $\Sigma_0$ is the infinite
volume chiral condensate at $\mu=0$.

For $N_t=4$ fig. 1 is an example of how the method works.
The points are the exact gap equation solution for the critical line
(both for real and imaginary $y$)
and a two parameter quadratic fit is satisfactory (upper continuous line).
The analytical continuation to real $y$ gives the decreasing
continuous line and its intersection with 
the $x^2-y^2=1$ line allows to determine the critical point.
Fig. 2 summarizes the results in the $(\beta,\mu)$ plane
compared with the exact ones as well as with the imaginary
chemical potential results based on a quadratic approximation
of the critical line.

\begin{figure}[t]
\vspace{9pt}
\centerline{\includegraphics*[width=2.3in,angle=270]
{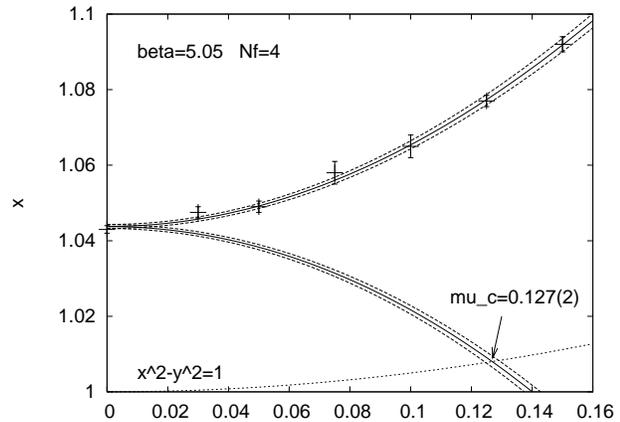}}
\caption{As in fig. 1: determination of the critical $\mu$ for 
$N_f=4$ QCD (the symbols are the critical points at 
imaginary $y$).}
\label{fig4}
\end{figure}
We get similar results in the $(\beta,y)$ plane: the results
are reported in fig. 3 and compared with the $(x,y)$
approach for the $N_t=8$ case .
Results from $(x,iy)$ plane are closer to the exact solution
respect to $(\beta,iy)$ ones and, when comparing with the 
standard imaginary chemical 
potential approach (considering in both cases an analytical 
continuation based on a quadratic two parameter fit), we see a clear
improvement when the critical chemical potential is large (see fig. 2).
Indeed we pick up the correct result up to $\mu/T\sim 8$.
From considering different $N_t$ values we realize that 
the quality of our results seems to decrease increasing $N_t$
and/or increasing the critical value of the chemical potential.
We considered up to $N_t=12$ (far beyond any optimistic prevision
for large scale simulations in finite density QCD in the next
years) and, up to this value, the quantitative agreement is good.

We have the first results for $N_f=4$ full QCD in a $8^3\times 4$ 
lattice at $m_q=0.06$. The critical point is located from
peaks in the susceptibility of plaquette, Polyakov loop or
chiral condensate (they all agree).
Fig. 4 shows the determination of the 
critical $\mu$ based on continuation to real $y$ of a quadratic
fit to the critical line in the $(x,iy)$ plane at $\beta=5.05$.
A consistent result is obtained from an independent analysis
carried out from a different data set based on numerical
simulations at fixed $\mu$ to determine the $(\beta,iy)$ 
critical line.

\end{document}